\documentstyle[11pt,epsf]{article}  %[aps]{revtex}

\begin{document}

%\draft
%\textit{}
\title{\bf Casimir Effect - The Classical Limit}
\author{M. Revzen and  A. Mann \\ \\
{\em Department of Physics, Technion -- Israel Institute of Technology,
Haifa 32000, Israel}} 

\date{}
\maketitle

\begin{abstract}
The temperature dependence of the Casimir effect for the radiation field 
confined between two conducting plates is analysed. The Casimir energy
is shown to decline exponentially with temperature while the Casimir
entropy which is defined in the text is shown to approach a limit
which depends only on the geometry of the constraining plates. The results are 
discussed in terms of the relation between the Bose distribution function
and the equipartition theorem - a relation based on a study by Einstein 
and Stern circa 1913. 
\end{abstract}

\newpage
%\vskip0.1cm
\section{Introduction}

The Casimir effect involves the difference between the energy of a field 
subject
to some constraining boundaries and the energy of the field with such
constraints removed. Although Casimir \cite{casimir} introduced this 
consideration long ago, in 1948, the 
study of this effect enjoys currently considerable popularity
 \cite{milonni,trunov,levin,pop}. Indeed
this journal included very recenty an exposition on the effect  \cite{amj}.
The experimental status of the effect was tenuous until 
 1997  with the
report of vindication of the theoretical prediction to an accuracy of a few 
percent
\cite{prl}. The reason for this continuous interest is not hard to find:
in this effect we have a direct demonstration of the existence of the so
called ``vacuum fluctuations'' or ``zero point energy''. Thus the original
effect was a prediction that two parallel conducting plates will
experience mutual attraction in vacuum (at zero temperature, $T = 0$), 
caused by modification of
the allowed modes due to the presence of the plates. Remarkably the attractive
force is independent of the coupling of the electromagnetic (EM) field to
matter ( viz., the electronic charge, e) and is proportional to the 
velocity of 
light, c and Planck's constant, h. Today, of course, we understand
that every quantized field exhibits fluctuations even at its lowest
(vacuum) state and these fluctuations sample the allowed modes. Furthermore, 
whereas the original effect related to
modified boundary conditions, more recent works (e.g., Schwinger as cited in
\cite{carlson}) led to studies of bulk (volume) Casimir effect (where we
consider the forces on a sample because it is different from its environment, 
 e.g., 
through having a different dielectric constant) and to the dynamical Casimir
effect where time is involved.\\ 

Another direction for Casimir effect
studies is its extensions to finite temperatures. Again the simplest case 
deals with 
modified modes of the EM field due to the presence of conductors as
boundaries but  now each mode is thermally occupied in addition to it being 
sampled through  vacuum
fluctuations. This also leads to an effective force between the plates.
Finite temperature studies were carried out by several authors
\cite{takagi}. Yet, to our knowledge, the classical limit of the effect,
even for the EM field, has not been clarified todate. This paper 
addresses this problem.\\

The paper is organized as follows. The next section, section 2, contains a 
review - hopefully self contained - of the Casimir effect problem in its 
simplest form, viz., radiation field confined between two conducting plates. 
This section includes the exact 
solution for the Casimir energy, free energy and entropy \cite{opher}. 
These are displayed in a figure which gives the  behavior of these quantities
for all temperatures. 
The succeeding section, Section 3, includes the central point of the paper, 
viz., an exposition of the classical limit and a study of its implications. 
Section 4 is devoted to the demonstration that the classical solution is 
``robust'' - by this we mean that a naive high temperature expansion for the 
Casimir energy doesn't have any nonvanishing corrections to 
the classical value. In the conclusions which are given in Section 5 we 
review the old (cf. \cite{milonni}) argument for the existence of zero 
point fluctuations as viewed from our vantage point.\\

\newpage

\section{The Casimir Effect}

Evaluation and definition of the Casimir energy at zero temperature due to 
vacuum fluctuations for the case under study here is given in several texts 
and reviews (e.g.,  \cite{milonni, trunov, greiner, zuber}). Our presentation 
therefore, although aspiring for self containment, is somewhat sketchy. 
We consider the radiation field confined between two conducting plates. 
The size of the plates edge is L. The first plate is placed at $z = 0$ in the 
$XY$
 plane, and the second at $ z = d $ parallel to the $XY$ plane. $L >> d$. 
 (In fact we are interested in $L \rightarrow \infty$ while $d$ remains 
finite.)
 The energy, i.e. the expectation value of the Hamiltonian,
 tied down in zero point fluctuations in the mode $k$ is,
 in obvious notation 
($k$ includes the polarization and the zero in the argument relates to 
$T = 0$),
\begin{equation}
E_k(0) = \frac{1}{2}\hbar \omega_k = \frac{1}{2} \hbar c|k|.
\end{equation}

The total energy density of the EM field with the conducting plates as
boundaries is ($k_{||}$ is the magnitude of the wave vector parallel to 
the plates \cite{zuber})
\begin{equation}
\frac{E(d, T=0)}{L^{2}d} = \frac{\sum_{k}E_{k}}{L^{2}d} = 
\frac{\hbar c}{2\pi d}\int^{\infty}_0 k_{||}dk_{||}\left[ \frac{k_{||}}{2} + 
\sum_{m=1}^{\infty}|k_m|\right]
\end{equation}
$$ k_{||}^2 = k_{x}^2 + k_{y}^2; \;\; k_{m}^2 = k_{||}^2 +
 \frac{m^2 \pi^2}{d^2}; \;\; m = 0,1,2,...$$

The energy density, in dimensionless units, $\varepsilon(d,0)$, is given by
 \cite{opher}
\begin{equation}
{E(d,0)\over L^2 d} = {\hbar c\over 2\pi^2}{\pi^4\over d^4}\varepsilon(d,0)
\equiv D\varepsilon(d,0);
\end{equation}
\begin{equation}
\varepsilon(d,0) = \int^{\infty}_0 xdx\ \left[\frac{x}{2} + \sum_{m=1}^{\infty}
\sqrt{x^2 + m^2}\right].
\end{equation}
The equation above defines D whose dimension is energy density. 
The energy density due to vacuum fluctuations of the radiation field in an 
arbitrarily large volume, $V$ (e.g., $V = L^3$) - which serves as the 
reference, unconstrained, system is
\begin{equation}
\frac{E(\infty,0)}{V} = \frac{1}{V}\sum_k \frac{1}{2}\hbar 
\omega_k = D \varepsilon(\infty,0),
\end{equation}
$$\varepsilon(\infty,0) = \int xdx\ \int dm\sqrt{x^2 + m^2},$$
The Casimir energy density, at $T=0$, in dimensionless units, is given by
\begin{equation}
\varepsilon_{c}(0) = \varepsilon(d,0) - \varepsilon(\infty,0).
\end{equation}                                                              
Both $\varepsilon(d,0)$ and $\varepsilon(\infty,0)$ diverge. $\varepsilon_c$ 
is commonly \cite{casimir, milonni, greiner} evaluated by a  physically 
justifiable regularization
technique. Thus a wave vector dependent function, $r(k/k_c)$, is introduced
into the above integrals 
such that for $k >> k_c$ $r \rightarrow 0$ while $r \rightarrow 1$ for 
$k << k_c$ thereby rendering the integrals convergent. A simple choice is 
($\alpha = 1/k_c$), $$ r(\alpha  k) = exp \left[-\alpha k \right].$$
$\alpha$ is allowed to go to zero at the end of the calculations - for all 
the terms together - thus the sum is ``regularized''. The details of the 
calculations will not be given here (cf. \cite {greiner, zuber}). This 
issue is further discussed in section 4. The result for our case is
\begin{equation}
\varepsilon_{c} (0) = -\frac{4}{(2\pi)^4} \zeta(4),
\end{equation}
where
$$ \zeta (n) = \sum_{m=1}^{\infty}\frac{1}{m^n}. $$
The force per unit area, ${\cal F}/L^2$,  between the plates is now calculable from the 
regularized Casimir energy, Eq. 7 , yielding, after reverting to physical 
dimensionality,
\begin{equation}
\frac{{\cal F}}{L^2} = - \frac{\delta E_c}{L^2 \delta d} = - 
\frac{\delta}{L^2 \delta 
d}\left[ E(d,0) - E(\infty,0)\right] = - \frac{\pi\hbar c}{240 d^4}.
\end{equation}
thereby yielding the well known Casimir force \cite{casimir}.\\

The finite temperature problem is quite similar. Now the zero point energy is
supplemented by the thermal energy. Thus the (average) energy tied down in 
the mode labeled by k is
\begin{equation}
E_k (T) = (1/2)\hbar \omega_k + \frac{\hbar \omega_k}{exp(\beta \hbar
 \omega_k) - 1}
 = \frac{\hbar \omega_k}{2}{\rm coth}\left[\frac{\beta \hbar \omega_k}{2}\right].
\end{equation}
Correspondingly,
\begin{equation}
\frac{E(d,T)}{L^2 d} = \frac{\hbar c}{2\pi d}\int k_{||}dk_{||}
\left[\frac{|k_{||}|}{2}{\rm coth}(\frac{\beta \hbar \omega(k_{||})}{2}) 
+ \sum_{m=1}^{\infty}|k_m|{\rm coth}(\frac{\beta \hbar \omega(k_m)}{2})\right].
\end{equation}
Returning to our dimensionless units we may write the energy density as a sum 
of a zero temperature part plus a temperature dependent part:
\begin{equation}
\varepsilon(d,T) = \varepsilon(d,0) + u'(d,T),
\end{equation}
\begin{equation}
u'(d,T) = \frac{f(0)}{2} + \sum_{m=1}^{\infty}f(m),
\end{equation}
$$ f(m) = \int\limits_{m}^{\infty}dy\ {y}^2 n(y,T),$$
\begin{equation}
n(y,T) = \frac{1}{\exp(\frac{T_c}{T}y) - 1}, \; \; \; k{_B}T{_c} = 
\hbar c\frac{\pi}{d}.
\end{equation}
The corresponding expressions for the unconstrained system are
\begin{equation}
\varepsilon(\infty,T) = \varepsilon(\infty,0) +u'(\infty,T),
\end{equation}
with $$ u'(\infty,T) = \int\limits_{0}^{\infty}dm\ f(m).$$
Evaluating the sum in Eq. 12 via the Poisson summation formula \cite{greiner, opher} gives ($\mu = 2\pi m$)
\begin{equation} 
\varepsilon_c(T) \equiv \varepsilon(d,T) - \varepsilon(\infty,T) = 
-4t^3\sum_{m=1}^{\infty}\frac{1}{\mu}{\rm coth}(t\mu){\rm csch}^2(t
\mu),\; \; \; t = \pi \frac{T}{T_c}.
\end{equation}

For $t << 1, \;\;\varepsilon_c \rightarrow \varepsilon_c(0)$.\\

The finite temperature Casimir energy, 
$\varepsilon_c(t)$, i.e. the expectation value of the Hamiltonian of the 
constrained system with the that of the unconstrained subtracted from it,
 is displayed in Fig.1. The relevant free energy was 
calculated by several authors \cite{takagi}. We outline the procedure as 
follows: The partition function (for one mode) is 
\begin{equation}
Z_k = \sum_{n=1}^{\infty}\exp\left[-\beta\hbar \omega_k(n +1/2)\right] = 
\frac{\exp\left[-\beta\hbar \omega_k/2\right]}{1 - \exp\left[-\beta\hbar 
\omega_k\right]},
\end{equation}
and hence the free energy for the mode labeled by k is
\begin{equation}
F_k = - k_B T ln Z_k = \frac{1}{2}\hbar \omega_k + k_B T \ln(1 -\ \exp\left[-\beta\hbar \omega_k\right]),
\end{equation}
Hence the expression for the free energy density of the constrained system is,
\begin{equation}
\frac{F(d,T)}{L^2 d} = \frac{E(d,0)}{L^2 d} + k_{B}T\frac{\hbar c}{2\pi 
d}\int k_{||}dk_{||}\left[ \ln(1 -e^{-\beta \hbar c k_{||}}) + 
2\sum_{m=1}^{\infty} \ln(1 - e^{-\beta \hbar c k_{m}})\right].
\end{equation}
A  corresponding equation holds for the unconstrained 
system thereby leading to the 
dimensionless expression for the Casimir free energy density, $ \phi_c $,
 given by
\begin{equation}
\frac{F_c(T)}{L^2 d} = D \phi_c(t),
\end{equation}
where D is defined by Eq. 3, and (detailed derivation is given in the appendix),
\begin{equation}
\phi_c(t) = -2t\sum_{m=1}^{\infty} {\frac{1}{\mu^3}}\left[{\rm coth}
(t\mu) +
 (t\mu){\rm csch}^2(t\mu)\right] .
\end{equation}
The Casimir free energy,
$\phi_c(t)$, is displayed in Fig. 1. For $t << 1$, $\phi_c(t) \rightarrow 
\varepsilon_c(0)$, as it should.

At this juncture it is natural to consider Casimir's Entropy 
\cite{opher}. This entropy, $\sigma_c(t)$, in our dimensionless units is defined by
\begin{equation}
\phi_c(t) = \varepsilon_c(t) - t\sigma_c(t).
\end{equation}
$ -\sigma_c(t)$ is displayed in Fig. 1 (note the negative sign). At high 
temperatures, $t >> 1$, $\varepsilon_c(t)$ 
falls off exponentially with $t$, $\sigma_c(t)$ approaches a constant value  
(independent of t and, of course, 
of $\hbar$) while $\phi_c(t)$ becomes proportional to $t$. Reexpressing 
these results in standard physical dimensions we get, at this limit, $t >> 1$,
\begin{equation}
E_c \rightarrow 0, \; \; \; F_c \rightarrow -TS_c, \; \; \; S_c \rightarrow \frac{\zeta(3)}{2^3 \pi}(\frac{L^2}{d^2}).
\end{equation}
Note that the entropy is proportional to the area of the plates scaled by $d^2$
(d is the plates' separation). We thus conclude that in the classical 
limit (high temperatures) the Casimir force is purely entropic. 
\newpage
%%%%%%%%%%%%%%%%%%%
\vspace*{-6cm}
\begin{figure}[htbp]
\epsfxsize=0.8\textwidth
%\centerline{\epsffile{/usr/users/revzen/casfig2.ps}}
\centerline{\epsffile{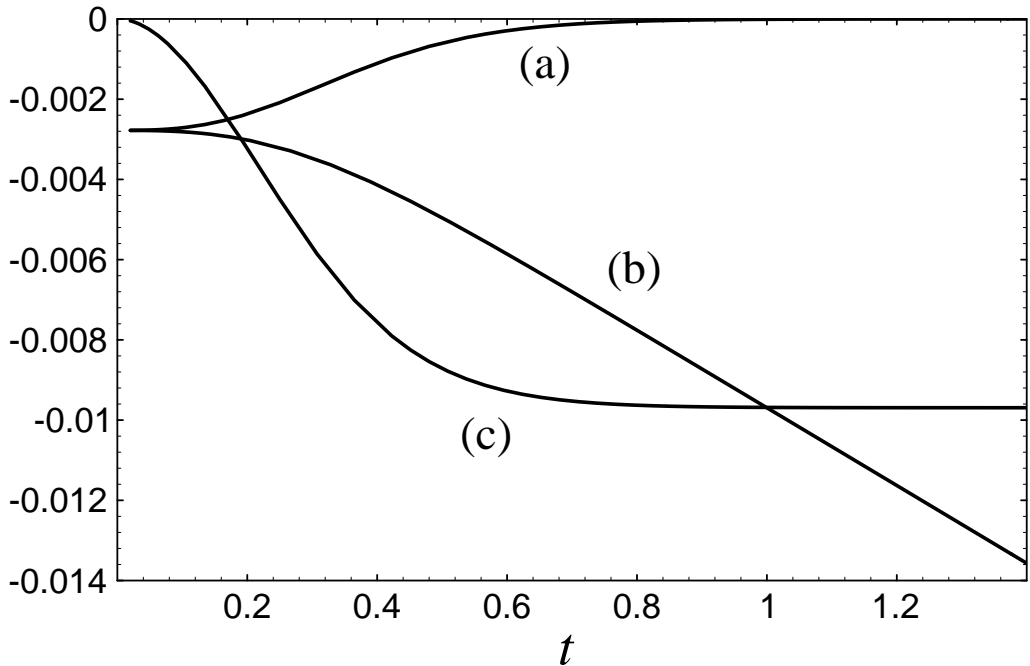}}
\vspace*{-4.5cm}
\caption{Casimir's Energy $\varepsilon_c$ (a), 
Free Energy $\phi_c$ (b), and 
Entropy $-\sigma_c$ (c) as a function of temperature, $t$.}
\end{figure}
%%%%%%%%%%%%%%%%%%%

\section{ The Classical Limit - Discussion}

An intuitive understanding of the vanishing of the Casimir energy for
 $T >> T_c$ ( cf. Fig. 1) may be gained via the following reasoning.
 Loosely speaking \cite{amj} the number of normal modes (per unit volume) 
in our
 confined system is unchanged upon changing $d$ (in the 
case under study). (A proof of this, based on regularization, is given in 
the next section.)   
Thus moving the walls adiabatically leads to shifts
 in the levels - not to appearance (or disappearance) of levels (modes).
 This (coupled with the classically valid equipartitioning of energy,
 which means that each harmonic oscillator-
like mode holds $k_BT$ amount of energy) implies that the energy density,
 in this classical limit, is unchanged upon   changing $d$. i.e., in the 
classical limit, ${\it defined}$ by the validity of the Rayleigh Jeans (RJ) 
law or the equipartioning of the energy \cite{revzen}, the Casimir 
energy is nil. This, since it is defined as the difference
 in the energy (density) between the constrained ($d << L$) and the 
unconstrained ($d \sim L$) cases and we have just argued that there exists 
(intuitive) one to one correspondence 
between the levels regardless of the size of $d$.   
Thus we have, in conformity with the result of the exact calculations as
 exhibited in Fig. 1, that,
\begin{equation}
\lim_{\frac{T}{T_c} \rightarrow \infty} \varepsilon_c(d,T) = 0.
\end{equation}  
In discussions of black body radiation, Kirchhoff's law (that the ratio of
 emissivity to absorptivity of all bodies is a universal function of the 
wavelengh, $\lambda$ and the temperature, $T$) is used \cite{milonni}
 to infer that $U$, the total electromagnetic ($EM$) energy in a cavity at 
thermal equilibrium is a function of 
$T$ only, $U = U(T)$ with $U$ the thermodynamic internal energy. We see
 from our discussion above that this is true only for $T >> T_c$ where 
$T_c$ is a  characteristic temperature of 
the cavity (cf. Eq. 13). (e.g., at $T = 0$ it is, strictly speaking, never 
true.)
 For this reason we refer to Eq. 23  as ``Kirchhoff's theorem''. We remark
 that the RJ result can be readily obtained  within classical physics 
\cite{revzen} and hence our definition of classical limit as the
 one where the equipartition theorem holds is a reasonable one.\\

The entropy, $S$, as a function of the energy, $E$, of a cavity 
(i.e., a constrained system), is given by (e.g., \cite{huang})
\begin{equation} 
S(E) = - k_B {\rm ln}\;\Sigma(E),
\end{equation}
where $\Sigma(E)$ is the number of states with energies less than or
 equal to E. In evaluating the Casimir entropy in the classical limit
 (which is the difference in the entropies, at high energies, 
 of the entropies of the constrained and the ``free'' systems) 
we expect the dominant contribution, the ``volume term''
 \cite{carlson}, to cancel. This is because at these energies
 the dominant contribution is from the short wavelength modes
 ($\lambda \sim  1/T$). These are insensitive to the boundaries when the
 dimension of the cavity much exceeds these wavelengths \cite{peierles}.
 The contribution to the Casimir entropy, then, is essentially 
determined by the  
long wavelengths which are geometry dependent, i.e.,  which relate to the 
``shape'' of the cavity and is independent of temperature. It follows 
then that the Casimir {\it free} energy is proportional to the temperature.
This reasoning
 is born out by the explicit calculational results. Hence we see that 
 this, viz., the {\it free} energy being proportional to T, is not at all 
related to the RJ law - the latter relates to 
the energy.\\

We are now in a position to interpret the zero point energy (zpe) as a
 contribution 
necessary to assure that at high temperatures , the energy, $U$, is a 
function 
of $T$ only as, indeed,  was noted long ago (1913) by Einstein and Stern 
(\cite{milonni}, p.2): without it the energy will depend on the boundaries.  
Alternatively, if we assume the validity of 
``Kirchhoff's theorem'' at 
high temperatures, we may deduce the zero temperature Casimir energy as 
follows.\\

Let the energy per mode, $k$, of the radiation field be written as ($T_c$ 
refers to the constraints, if present)       
\begin{equation}
u(k,T;T_c) = u(k,0;T_c) + u'(k,T;T_c)
\end{equation}
where $u'$ is the energy held in the mode without the zpe. 
For an allowed mode, $k$:
\begin{equation}
u'(k,T;T_c) = \frac{\hbar \omega_k}{\exp(\beta \hbar \omega_k) - 1}.
\end{equation}
The Casimir energy is then,
\begin{equation}
\varepsilon_c(T;T_c) = \sum_{k} u(k,T;T_c) - \sum_{k} u(k,T).
\end{equation}
Use of  Eq. 25 ( i.e., separating the thermal energy from the vacuum's) and 
Kirchhoff's theorem ( Eq. 23) implies,
\begin{equation}
\lim_{T/{T_c} \rightarrow \infty} {\left[ \sum_{k} u'(k,T;T_c) - 
\sum_{k} u'(k,T) +  \varepsilon_{c}(0)\right]} = 0.
\end{equation}
with $\varepsilon_c(0)$ being the Casimir energy at zero temperature. Hence
\begin{equation}
\lim_{T/{T_c} \rightarrow \infty} \left[ \sum_{k} u'(k,T;T_c) -
 \sum_{k} u'(k,T) \right] = -  \varepsilon_c(0).
\end{equation}
Now both terms in the square bracket are readily calculable - their 
temperature dependence assures convergence - and thus yield the Casimir 
energy at $T=0$ (  $\varepsilon_c(0)$ )  without recourse to a (perhaps) 
objectionable regularization scheme.\\

%\newpage

\section{Robustness of the Classical Limit}

In this section we evaluate the Casimir energy through a power series 
expansion in $t$ ($t \equiv  \pi \frac{T}{T_c}$). This scheme requires 
regularization for each term in the expansion. The result is that, within 
such an expansion, Kirchhoff's theorem is exact. i.e. the Casimir energy 
vanishes to all orders in this high temperature expansion.
This is interpreted as implying that the classical equipartition theorem is
robust -  the classical approximation, once taken, is exact to within 
power series corrections. We hasten to add that this is yet another example 
of incorrect handling of infinities which are further discussed at the end 
of this section.\\

Let us return to the expression for the Casimir energy density at finite 
temperature, Eqs. 10 and 15, which, in our dimensionless units are summarized 
by 
$(x_m = \sqrt{x^2 + m^2})$,
\begin{equation}
\varepsilon_c(t) = \int\limits_{0}^{\infty}xdx 
\left[\frac{x}{2}{\rm coth}(\frac{\pi x}{2t}) + \sum_{m=1}^{\infty}x_m 
{\rm coth}(\frac{\pi x_m}{2t}) - \int\limits_{0}^{\infty}dm\;{\rm coth}
(\frac{\pi x_m}{2t})\right].
\end{equation}
We now adjunct to each of the above integrals the ``standard'' 
\cite{greiner,zuber} cutoff function, 
$$ f =exp\left[ -\alpha x_m \right], $$ 
assuring thereby the convergence of the integrals. ( We are interested in the 
$\alpha \rightarrow 0$ limit.)
We note that such regularization scheme is justifiable on physical grounds as 
follows \cite {greiner, zuber, trunov}. The conductivity is, in general, 
a function of frequency. Indeed, all metallic conductors are effectively 
transparent to radiation with wave lengths comparable to the interatomic 
spacing,
 $a$. Hence the integrals (and summations) over wavenumbers,
 $k$, are limited to 
$k \leq k_c \sim 1/a$, i.e., for wavenumbers $|k| > k_c$ the plates do not 
provide a boundary and hence the Casimir energy for $k > k_c$ evidently 
vanishes.\\
Assuming the validity of the ``standard'' regularization scheme (each 
integrand is multiplied by a cutoff function assuring the convergence 
of the integrals)
 we may expand \cite{abramowitz}, p.35,
\begin{equation}
{\rm coth} z = \frac{1}{z} + \frac{z}{3} - \frac{z^3}{45}.\ ...\frac{2^{2n}B_{2n}}{(2n)!}z^{2n-1}; \; \; \; B_{2n}\ are\ Bernouli\ numbers.
\end{equation}
The evaluation of $\varepsilon_c(t)$ reduces to the evaluation of terms like 
\cite{opher}
\begin{equation}
g_p = \int_{0}^{\infty}xdx\ \frac{x^p}{2}f(x) + 
\sum_{m=1}^{\infty}\int_{m}^{\infty}xdx\ x^pf(x) - 
\int_{0}^{\infty}dm \int_{m}^{\infty}xdx\ x^pf(x) .
\end{equation} 
The cutoff parameter, $\alpha$, is set equal to zero at the end of 
the calculations;
the result is
$$ g_p = \lim_{\alpha \rightarrow 0}\frac{ d^{p+1} }{d\alpha^{p+1}}
\left[ \frac{1}{2\alpha} + \frac{1}{ \alpha^2 } 
\frac{\alpha}{ exp(\alpha)-1 } - \frac{1}{ \alpha^2 } \right] . $$
Noting that \cite{abramowitz}, p.1076,
$$ \frac{y}{exp(y)-1} = \sum_{n=1}^{\infty}\frac{B_{n}y^{n}}{(n)!},$$
we get $g_p = 0$ for $p$ even, and for p odd, 
$$ g_p = \frac{B_{p+1}}{(p+3)(p+2)} \;\;\; for\ p = n - 3. $$
Returning to our expression, Eq. 30, we see that upon substituting 
Eq. 31  only terms with $p$ even occur, i.e., 
$\varepsilon_c(T) = 0$. This confirms the exponential decay of the 
Casimir energy with temperature that is implied by Eq. 15, and depicted in 
Fig. 1. (The case of $p = 0$ will be recognized as the sum over
 the constrained number 
of modes, per unit volume, with the unconstrained number deducted therefrom. 
That the result is nil constitues a proof, based on a particular 
regularization scheme, of our 
assersion (Section 3) that the number of modes, per unit volume, is unchanged 
upon varying the plates' separation, d.)\\ 
 
This ``robustness'' of the classical (erroneous) solution is worthy of note 
and it implies that the correct expression can't be obtained from the wrong 
one by analytic (power series) means.

%\newpage

\section{Concluding Remarks}

This paper gives the classical - here the high temperature - limit of the 
Casimir effect. The case studied was the simplest and historically the first 
considered - viz. the radiation field confined between two large 
(in the limit - 
infinite) parallel conducting plates separated by a relatively short distance,
d. We argued that  in the classical limit (defined to be temperatures such 
that
the Rayleigh Jeans, i.e. energy equipartition, theorem holds) ``Kirchhoff's 
theorem'' is valid - i.e. the energy density of the radiation field is a
function of  the temperature only and this implies that, in this limit, the
Casimir
energy vanishes. We showed that the zero point energy is required to 
assure the validity of ``Kirchhoff's theorem'' in the classical limit. 
Alternatively, assuming the validity of the theorem allows the evaluation of 
the Casimir energy 
at $T = 0$ without recourse to any regularization scheme. 
We noted, following reference \cite{milonni}, that these results were 
anticipated by Einstein and Stern in 1913, prior to the formulation of 
quantum mechanics and quantum field theory.
Thus these authors noted that the high temperature expansion of what 
we now term Bose distribution function is ($\beta \hbar \omega << 1$)
$$ \frac{\hbar \omega}{\exp(\beta \hbar \omega) - 1} \rightarrow k_BT - 
\frac{1}{2}\hbar \omega + O(1/T).$$
Thus we have here a temperature independent term which contributes to 
the total energy. Its cancellation, i.e., the validity of 
``Kirchhoff's theorem'' in our presentation, requires a positive zero 
point energy, $+ \frac{1}{2}\hbar \omega$. Thus the removal of the zero 
point energy by considering a ``normally'' ordered Hamiltonian does not 
eliminate the need for zero point energy. It (the zero point energy) is 
seen, ala the Einstein and Stern result as discussed above, to be required
for the correct classical limit or, in our terminology, for the validity
of ``Kirchhoff's theorem''. This stems here from the form of the Bose
distribution rather than the Hamiltonian.\\

The evaluation of the (Casimir) free energy in the classical limit led to 
showing that the Casimir entropy, which is defined in a natural way 
in the text, is (in this limit) temperature independent constant and reflects 
finite volume corrections in statistical physics. Thereby
the proportionality of the free energy to the temperature, 
$T$ that is well known \cite{prl} is seen not to be related to the Rayleigh 
Jeans limit \cite{boyer} but rather is a consequence of the perhaps 
interesting result  that the Casimir force in the classical 
limit is purely entropic - in fact geometric. We have
demonstrated that the classical limit results, i.e., at temperatures high 
enough 
to warrant the applicability of the Rayleigh Jean's law  are ``robust'' in 
the sense that (naively) it leads to vanishing corrections.\\

\newpage

\vskip0.4cm

{\bf Acknowledgements:}

\noindent This work was supported by GIF -- German--Israeli Foundation for
Research and Development, the Fund for promotion of Research at the 
Technion, and the Technion VPR Fund - Promotion of Sponsored Research.  
Special thanks are due to our colleagues Constantin Brif, 
Hiroshi Ezawa, Koichi Nakamura, Lev Pitaevski, Giuseppe Vitiello  and Joshua 
Zak for informative comments.
\vskip0.7cm

{\bf Appendix}: Derivation of Equation (20).\\

For $k_{||} = \frac{\pi}{d}x,\; \eta = T_c/T$  the expression for 
$\frac {F(d,T)}{L^2 d}\;- \frac{E(d,0)}{L^2 d}$ is, $$k_B T \frac{1}{\pi ^2}
(\frac {\pi}{d})^3 \left[\int\frac{1}{2} x dx \ln(1 - e^{-\eta x}) +
\sum_m \int _{m}^{\infty}x dx \ln(1 - e^{-\eta x})\right].$$
Defining $$F(m) = \int _{m}^{\infty} x dx \ln(1 - e^{-\eta x}),$$ 
the Poisson summation formula reads  ($\mu \equiv 2\pi m$),
$$ \frac{1}{2}F(0) + \sum _{m=1}^{\infty}F(m) = \int _{0}^{\infty}F(x) dx +
2\sum _{m=1}^{\infty}\int _{0}^{\infty} \cos(\mu x) dx
\int_{x}^{\infty} y dy
\ln(1 - e^{-\eta y}).$$
Noting that 
$$\int x^2 ln(1 - e^{-\eta x})dx = \frac{2}{\eta}\zeta(4),$$
 integration by parts yields (\cite{abramowitz} p.584),
$$2 \sum_{m=1}^{\infty}\int \cos(\mu x)dx \int_{x}^{\infty} y dy
\ln(1-e^{-\eta y}) = 2\sum \left[\frac{\eta}{\mu^4}- \frac{\pi}{2\mu^3}
\coth(\frac {\pi\mu}{\eta}) - \frac{\pi}{2\mu^2} {\rm csch}^2(\frac{\pi\mu}
{\eta}) \frac{\pi}{\eta}
\right].$$
Substracting $\frac{F(\infty,T)}{L^3}$ gives Equation 20.

\end{document}